\documentclass[12pt,english]{article}
\usepackage{latexsym}
\usepackage{amssymb}
\usepackage{epsfig}
\usepackage{fancyhdr}
\usepackage{float}
\usepackage[T1]{fontenc}
\usepackage{graphics}
\usepackage{graphicx}
\usepackage[latin1]{inputenc}
\usepackage{multicol}
\usepackage{pslatex}
\usepackage{pifont}
\usepackage{rotating}
\usepackage{supertabular}
\usepackage{t1enc}
\usepackage{units}
\usepackage{varioref}
\usepackage{wrapfig}
\usepackage{subfigure}
\usepackage{lscape}
\usepackage{slashbox}
\usepackage[active]{srcltx}
\def\0{\phantom{0}}

\begin{document}

\pagenumbering{arabic}
\baselineskip25pt

\begin{center}
% Title
{\bf \large Unlike Lennard-Jones Parameters for Vapor-Liquid Equilibria}\\
\bigskip

% Authors
\renewcommand{\thefootnote}{\fnsymbol{footnote}}
 Thorsten Schnabel, Jadran Vrabec\footnote{author to whom correspondence should be addressed, Tel.: +49-711/685-66107, Fax:
+49-711/685-66140, Email: vrabec@itt.uni-stuttgart.de}, Hans Hasse \\
\renewcommand{\thefootnote}{\arabic{footnote}}

% Address
Institut f\"ur Technische Thermodynamik und Thermische Verfahrenstechnik, \\
Universit\"at Stuttgart, D-70550 Stuttgart, Germany \\
\end{center}

% Abstract
\begin{abstract}
\baselineskip25pt

\noindent 
The influence of the unlike Lennard-Jones (LJ) parameters on vapor-liquid equilibria of mixtures is investigated and the performance of eleven combining rules is assessed. In the first part of the work, the influence of the unlike LJ size and energy parameter on vapor pressure, bubble density and dew point composition is systematically studied for the mixtures CO+$\rm C_2H_6$ and $\rm N_2$+$\rm C_3H_6$, respectively. It is found that mixture vapor pressure depends strongly both on the size and the energy parameter whereas the bubble density depends mostly on the size parameter and the dew point composition is rather insensitive to both parameters. In preceding work, unlike LJ parameters were adjusted to experimental binary vapor-liquid equilibria for 44 real mixtures. On the basis of these results, in the second part of the work eleven combining rules are assessed regarding their predictive power. A comparison with the adjusted unlike LJ parameters determined from the fit shows that none of the eleven combining rules yields appropriate parameters in general. To obtain an accurate mixture model, the unlike dispersive interaction should therefore be adjusted to experimental binary data. The results from the present work indicate that it is sufficient to use the Lorenz rule for the unlike LJ size parameter and to fit the unlike LJ energy parameter to the vapor pressure.

\end{abstract}


\clearpage

% Main text
\section{Introduction}

In molecular simulations of a binary mixture A+B with pairwise additive potentials, three different interactions occur: two like interactions between molecules of the same type A-A and B-B, which are fully defined by the pure component models, and the unlike interaction between molecules of different type A-B. In mixtures consisting of polar molecules, the electrostatic part of the unlike interaction is fully determined by the laws of electrostatics. However, there is no rigorous physical framework that yields reliable unlike repulsion and dispersion parameters like the Lennard-Jones (LJ) parameters studied in the present work. For finding these parameters, combining rules were developed in the past based on physical and mathematical intuition or on empirical approaches. Eleven of these combining rules were investigated in the present work. These combining rules rely soley on pure component data, namely the LJ parameters and, in some cases, additionally the polarizablility $\alpha$ or the  ionization potential $I$. Other combining rules, that are not discussed in this work, also employ dispersion force coefficients \cite{Pena1980,Tang1986,Bzowski1988}, diamagnetic susceptibility \cite{Mason1964,Good1971} or effective transition energies \cite{Pena1982,Pena1982a}. 
  
Another approach for obtaining unlike LJ parameters is to adjust them directly to experimental binary data, for which a single data point may in principle be sufficient. Kohler et al.~\cite{Kohler1982} fitted both unlike LJ parameters to experimental second virial coefficients of binary mixtures. Following this approach, M{\"o}ller et al.~\cite{Moeller1992} developed a method for adjusting the unlike LJ size and energy parameters to experimental excess volumes and enthalpies. Their investigations
showed that the unlike LJ size parameter determined from the fits practically does not deviate from the arithmetic mean of the pure component LJ size parameters. This finding is supported by the work of Vrabec and Fischer for $\rm Ar+CH_4$ \cite{Vrabec1995} as well as for the three binary mixtures that can be formed out of $\rm CH_4$, $\rm C_2H_6$ and $\rm CO_2$ \cite{Vrabec1996}. Kronome et al.~\cite{Kronome2000} applied a similar approach, specifying the unlike LJ size parameter by the arithmetic mean. They adjusted only the LJ energy parameter to the experimental excess Gibbs enthalpy for the mixture $\rm N_2$+$\rm C_2H_6$ and obtained favorable results for vapor-liquid equilibria.

In previous work of our group \cite{Stoll2003,Vrabec2004,Stoll2004}, unlike LJ energy parameters of 44 binary mixtures were fitted to one experimental binary vapor pressure each. These mixtures contain noble gases, homonuclear and heteronuclear diatomics, small hydrocarbons, carbon dioxide, carbon disulfide and halogenated hydrocarbons, i.e.~refrigerants.  Overall 22 components are studied, where the molecules are composed out of 1 to 9 atoms. These components were modeled by the one-center Lennard-Jones potential (1CLJ) \cite{Vrabec2001}, the symmetric two-center Lennard-Jones potential either with a pointdipole (2CLJD) \cite{Stoll2003a} or with a linear elongated pointquadrupole (2CLJQ) \cite{Vrabec2001}. The parameters of the pure component molecular models were adjusted to experimental pure component bubble density and vapor pressure data \cite{Vrabec2001,Stoll2003a}. The pure component models are accurate, the mean errors of the vapor pressure, bubble density and heat of vaporization are typically 4 \%, 1 \% and 3 \%, respecitvely. Vapor-liquid equilibria of the 44 mixtures are described with typical deviations in vapor pressure, bubble density and in dew point composition of below 5 \%, 1 \% and 0.02 mol/mol, respectively \cite{Stoll2003,Vrabec2004,Stoll2004}. These mixture models also predict favorably other fluid properties like Joule-Thomson inversion curves \cite{Vrabec2004a} or self-diffusion and binary Maxwell-Stefan diffusion coefficients as well as bulk viscosities and thermal conductivities \cite{Fernandez2004,Fernandez2004a}. 

Ternary mixtures are predicted without further parameterization since the force fields are based on pairwise additive interactions. Results for ternary vapor-liquid equilibra
of six mixtures \cite{Stoll2003,Vrabec2004,Stoll2004,Vrabec1997a} confirm the predictive power of such models. 

\section{Combining rules \label{coru}}

The LJ potential $u_{ij}^{\rm LJ}$ is the most widely used functional form for describing repulsion and dispersive attraction. This pairwise additive potential acts between two molecules $i$ and $j$ and is given by

\begin{equation}
        u_{ij}^{\rm LJ}\left( r_{ijab}\right) =\sum_{a=1}^{m1}\sum_{b=1}^{m2}4\epsilon_{ab}\left[ \left( \frac{\sigma_{ab}}{r_{ijab}}\right)^{12}-\left( \frac{\sigma_{ab}}{r_{ijab}}\right)^6\right]\, , \label{equ_LJ}
\end {equation}

\noindent where $a$ is the site index of molecule $i$ and $b$ the site index of molecule $j$. $m1$ and $m2$ are the number of LJ interaction sites of molecule $i$ and $j$, respectively. The site-site distance is $r_{ijab}$, $\sigma_{ab}$ and $\epsilon_{ab}$ are the LJ size and energy parameters between sites $a$ and $b$. For obtaining the LJ parameters of the unlike interaction, denoted by $\sigma_{ab}$ and $\epsilon_{ab}$ subsequently, usually combining rules are used.

\noindent \textbf{Lorentz-Berthelot (LB)}\\ %[1ex]
The Lorentz-Berthelot combining rule is most widely used to determine the unlike LJ parameters. Lorentz \cite{Lorentz1881} proposed to use the arithmetic mean for the unlike size parameter motivated by collisions of hard spheres

\begin{equation}
\sigma_{ab}^{\rm LB}
=\frac{\sigma_{aa}+\sigma_{bb}}{2}\,. \label{equ_LB_sig}
\end{equation}

Berthelot \cite{Berthelot1898} proposed with little physical argument the geometric mean for the unlike energy parameter

\begin{equation}
\epsilon_{ab}^{\rm LB}
=\sqrt{\epsilon_{aa}\epsilon_{bb}}\,. \label{equ_LB_eps}
\end{equation}

The LB combining rule is by far the oldest and most common approach, but due to the fact that it can lead to inaccurate mixture properties \cite{Delhommelle2001,Ungerer2004}, numerous other combining rules have been developed.

\noindent \textbf{Kohler (K)}\\ %[1ex]
Kohler \cite{Kohler1957} used the approach of London \cite{London1930,London1937} for the dispersion energy to derive a combining rule for the unlike LJ energy parameter which uses the polarizability $\alpha$

%\begin{equation}
%\epsilon_{ab}^{\rm K}= \left[\frac{2\sqrt{\sigma_{aa}\sigma_{bb}}}{\sigma_{aa}+\sigma_{bb}}\right]^6\frac{2\alpha_{aa}\alpha_{bb}\left(\sigma_{aa}\sigma_{bb}\right)^3}{\alpha_{bb}^2\sigma_{aa}^6\epsilon_{aa}+\alpha_{aa}^2\sigma_{bb}^6\epsilon_{bb}}\epsilon_{aa}\epsilon_{bb}\,.
%\label{equ_K_eps2}
%\end{equation}

\begin{equation}
\epsilon_{ab}^{\rm K}= 2^7\left[\frac{\sigma_{aa}\sigma_{bb}}{\sigma_{aa}+\sigma_{bb}}\right]^6 \frac{\alpha_{aa}\alpha_{bb}}{\alpha_{bb}^2\sigma_{aa}^6\epsilon_{aa}+\alpha_{aa}^2\sigma_{bb}^6\epsilon_{bb}}\epsilon_{aa}\epsilon_{bb}\,. \label{equ_K_eps2}
\end{equation}

The application of polarizability in combining rules is not straightforward. Generally, the polarizability quantifies the distortion of the overall charge distribution of an atom, group or molecule by an electric field. It is in fact a tensor, which includes the anisotropic contributions of the directionally distorted charge distribution. This tensor is often described by a scalar $\alpha$ which is appropriate to characterize the distortion sensitivity for the whole molecule. Laidig and Bader \cite{Laidig1990} suggest that atomic or group polarizabilities influenced by the chemical enviroment additively contribute to the overall molecular polarizability. Hence, the atomic or group polarizabilities required to apply the combining rules to the symmetric two-center LJ potential ascribe 50 \% of the molecular polarizability to each of the two LJ sites. For molecules modeled with the one-center LJ potential, e.g.~methane, the total molecular polarizability was used here. Polarizabilities were taken from \cite{Lide1995-1996} except for the refrigerants R125 \cite{Wood2002,Fermeglia2003}, R134a \cite{Wood2002,Fermeglia2003} and R152a \cite{Fermeglia2003,Echt1988}.

Furthermore, Kohler proposed to use the arithmetic mean for the unlike LJ size parameter (Lorentz rule) as defined by Equation (\ref{equ_LB_sig}).

\noindent \textbf{Hudson-McCoubrey (HMC)}\\ %[1ex]
Hudson and McCoubrey \cite{Hudson1960} reformulated Reed's proposal \cite{Reed1955a} for the calculation of unlike LJ energy parameters which considers the ionization potential $I$ of the pure components and volume effects of the mixture. They claim that their expression is ``suitable to obtain viral coefficients of binary gas mixtures and binary gas-liquid critical temperatures'' \cite{Hudson1960} 

\begin{equation}
\epsilon_{ab}^{\rm HMC} =\frac{2\sqrt{I_{aa}I_{bb}}}{I_{aa}+I_{bb}}\left[\frac{2\sqrt{\sigma_{aa}\sigma_{bb}}}{\sigma_{aa}+\sigma_{bb}}\right]^6\sqrt{\epsilon_{aa}\epsilon_{bb}}\,. \label{equ_HMC_eps}
\end{equation}

\noindent The ionization potential is defined as the energy which is necessary to remove the outermost electron from an atom or molecule. Since the outermost electron cannot be explicitly assigned to a specific atom or group of a molecule, the full molecular ionization potential is ascribed in the present work to every LJ site of the molecular model. Ionization potentials were taken from \cite{Lide1995-1996} except for the refrigerants R125 \cite{Wood2002,Fermeglia2003}, R134a \cite{Wood2002,Fermeglia2003} and R152a \cite{Fermeglia2003,Echt1988}.

Srivastava and Madan \cite{Srivastava1953} derived the same expression as Equation (\ref{equ_HMC_eps}) with an approach similar to that of Kohler \cite{Kohler1957}. Note that Equation (\ref{equ_K_eps2}) can be transformed to Equation (\ref{equ_HMC_eps}) if the following relation between polarizability, ionization potential, LJ size and energy parameters is assumed for both components $a$ and $b$

\begin{equation}
4\epsilon_{aa}\sigma_{aa}^6=\frac{3}{4}\alpha_{aa}^2 I_{aa}\,.
\label{equ_aIepssig}
\end{equation}

\noindent Some authors, e.g.~Good and Hope \cite{Good1971} or Reid and Leland \cite{Reid1965}, suggest that the ratio between the geometric and arithmetic mean of the ionization potentials in Equation (\ref{equ_HMC_eps}) can be set to unity. However, in the present work Equation (\ref{equ_HMC_eps}) was applied.

For the unlike LJ size parameter in the HMC combining rule the Lorentz rule is applied as defined by Equation (\ref{equ_LB_sig}).

\noindent \textbf{Fender-Halsey (FH)}\\ %[1ex]
Fender and Halsey \cite{Fender1962} proposed to use the harmonic mean for the unlike LJ energy parameter in their empirical combining rule 

\begin{equation}
\epsilon_{ab}^{\rm FH}=\frac{2\epsilon_{aa}\epsilon_{bb}}{\epsilon_{aa}+\epsilon_{bb}} \label{equ_FH_eps}\,,
\end{equation}

\noindent which always gives smaller or equal values than the geometric mean. Mixtures of noble gases have mostly smaller unlike LJ energy parameters than those obtained with the geometric mean. Hence, Fender and Halsey \cite{Fender1962} achieved with their harmonic mean combining rule more accurate second virial coefficents for the mixture argon+krypton than with the LB combining rule.

For the unlike LJ size parameter in the FH combining rule the Lorentz rule is applied as defined by Equation (\ref{equ_LB_sig}).

\noindent \textbf{Hiza (H)}\\[1ex]
Hiza et al.~\cite{Hiza1969,Hiza1970,Hiza1978} used a semi-empirical approach to correct the LB combing rule. The expressions for the correction factors, dependent on the ionization potentials and empirical factors, were determined by fitting to equilibrium, transport and beam scattering data of noble gases and light hydrocarbons

\begin{equation}
\sigma_{ab}^{\rm H}=\left(1+0.025\cdot k_{ab}\right)\frac{\sigma_{aa}+\sigma_{bb}}{2} \label{equ_H_sig1}\,,
\end{equation}

and

\begin{equation}
\epsilon_{ab}^{\rm H}=\left(1-0.18\cdot k_{ab}\right)\sqrt{\epsilon_{aa}\epsilon_{bb}} \label{equ_H_eps1}\,,
\end{equation}

with

\begin{equation}
k_{ab}=\sqrt{I_{aa}-I_{bb}}\,\ln\left(\frac{I_{aa}}{I_{bb}}\right) \label{equ_H_eps2}\,,
\end{equation}

\noindent where the component order has to be chosen so that $I_{aa}>I_{bb}$.

Since $k_{ab}$ from Equation (\ref{equ_H_eps2}) is always larger than zero for nonidentical ionization potentials, the resulting unlike LJ size parameters are always larger than those from the Lorentz rule. The opposite holds for the unlike LJ energy parameters when compared to the Berthelot rule.

\noindent \textbf{Sikora (S)}\\ %[1ex]
In Sikora's combining rule \cite{Sikora1970} repulsion is considered via deformation energies of the electron clouds and resulting unsymmetric collision diameters which occur at small distances between overlapping atoms or molecules

\begin{equation}
\epsilon_{ab}^{\rm S} = 2^{15} \frac{I_{aa}I_{bb}}{\left(I_{aa}+I_{bb}\right)^2}
\frac{\sqrt{\epsilon_{aa}\sigma_{aa}^{12}\epsilon_{bb}\sigma_{bb}^{12}}}{\left[ \left(\epsilon_{aa}\sigma_{aa}^{12}\right)^{\frac{1}{13}} + \left(\epsilon_{bb}\sigma_{bb}^{12}\right)^{\frac{1}{13}} \right]^{13}}
\sqrt{\epsilon_{aa}\epsilon_{bb}} \,. \label{equ_S_eps}
\end{equation}

Applying similar considerations for the unlike LJ size parameter \cite{Maitland1981}, the following equation was proposed

\begin{equation}
\sigma_{ab}^{\rm S}=2^{-\frac{13}{12}}\left(\sigma_{aa}^{\frac{12}{13}}+\sigma_{bb}^{\frac{12}{13}}\right)^{\frac{13}{12}}\,.
\label{equ_S_sig}
\end{equation}

\noindent \textbf{Smith-Kong (SK)}\\ %[1ex]
Smith \cite{Smith1972} developed a combining rule considering the repulsive interaction of simple molecules. His combining rule includes the atomic distortion theory for repulsion. Kong \cite{Kong1973} used Smith's approach and a geometric mean relationship for the attractive interactions and applied it to the LJ potential leading to

\begin{equation}
\sigma_{ab}^{\rm SK}
=\left\{\frac{\left[\left(\epsilon_{aa}\sigma_{aa}^{12}\right)^{\frac{1}{13}}+\left(\epsilon_{bb}\sigma_{bb}^{12}\right)^{\frac{1}{13}}\right]^{13}}{2^{13}\sqrt{\epsilon_{aa}\sigma_{aa}^{6}\epsilon_{bb}\sigma_{bb}^{6}}}\right\}^{\frac{1}{6}} \,,
\label{equ_SK_sig}
\end{equation}

and

\begin{equation}
\epsilon_{ab}^{\rm SK}
=\frac{2^{13}\epsilon_{aa}\sigma_{aa}^{6}\epsilon_{bb}\sigma_{bb}^{6}}{\left[\left(\epsilon_{aa}\sigma_{aa}^{12}\right)^{\frac{1}{13}}+\left(\epsilon_{bb}\sigma_{bb}^{12}\right)^{\frac{1}{13}}\right]^{13}}\,.
\label{equ_SK_eps}
\end{equation}

\noindent \textbf{Halgren (HHG)}\\ %[1ex]
From investigations of noble gas mixtures using ab initio and experimental data, Halgren \cite{Halgren1992} deduced that the experimental pattern of the van der Waals minimum energy distances is well described by the ``cubic-mean'' combining rule

\begin{equation}
\sigma_{ab}^{\rm HHG}=\frac{\sigma_{aa}^3+\sigma_{bb}^3}{\sigma_{aa}^2+\sigma_{bb}^2} \label{equ_HHG_sig}\,.
\end{equation}

The van der Waals minimum energy distance is associated with the LJ size parameter. To introduce the experimental unlike noble gas values into the LJ energy parameters, a combination of the geometric and harmonic mean was applied which is claimed to yield good descriptions of the experimental noble gas values

\begin{equation}
\epsilon_{ab}^{\rm HHG}=\frac{4\epsilon_{aa}\epsilon_{bb}}{\left(\sqrt{\epsilon_{aa}} + \sqrt{\epsilon_{bb}}\right)^2}\,. \label{equ_HHG_eps}
\end{equation}

\noindent \textbf{Waldman-Hagler (WH)}\\ %[1ex]
Waldman and Hagler \cite{Waldman1993} derived their combining rule by applying mathematical methods which consider symmetry, uniformal scaling and simplification of a two-parameter to a single-parameter problem to deduce the general functional form of their combining rule. From a ``graphical analysis'' of noble gas mixture data they obtained

\begin{equation}
\sigma_{ab}^{\rm WH}=\left[\frac{\sigma_{aa}^6+\sigma_{bb}^6}{2}\right]^{\frac{1}{6}} \label{equ_WH_sig}\,,
\end{equation}

and

\begin{equation}
\epsilon_{ab}^{\rm WH}=\frac{2\sigma_{aa}^3\sigma_{bb}^3}{\sigma_{aa}^6+\sigma_{bb}^6}\sqrt{\epsilon_{aa}\epsilon_{bb}}\,.
\label{equ_WH_eps}
\end{equation}

\noindent \textbf{Al-Matar (M1 and M2)}\\ %[1ex]
Al-Matar and Rockstraw \cite{Al-Matar2004} proposed combining rules which are based on similar mathematical considerations as used by Waldman-Hagler (see above). The first of their combining rule, abbreviated in the following with M1, was obtained by functional analysis of experimental noble gas mixture data. They consider the pure component LJ parameters equally weighted. M1 is given by

\begin{equation}
\sigma_{ab}^{\rm M1} =\frac{1}{3}\sum_{L=0}^2\left(\frac{0.25\left(\sigma_{aa}^3+\sigma_{bb}^3\right)^2}{\sigma_{aa}^L\sigma_{bb}^L}\right)^{\frac{1}{6-2L}}\,, \label{equ_M1_sig}
\end{equation}

and

\begin{equation}
\epsilon_{ab}^{\rm M1} =\frac{3\sigma_{aa}^3\sigma_{bb}^3}{\sum\limits_{L=0}^2\left(\frac{0.25\left(\sigma_{aa}^3+\sigma_{bb}^3\right)^2}{\sigma_{aa}^L\sigma_{bb}^L}\right)^{\frac{6}{6-2L}}} \sqrt{\epsilon_{aa}\epsilon_{bb}}\,.
\label{equ_M1_eps}
\end{equation}

The second combining rule, abbreviated in the following with M2, uses weighting matrices which can account for uneven contributions of the pure component LJ parameters to the unlike quantities. The weighting matrices of M2 were obtained by fitting them to noble gas mixture data. The unlike parameters are given by 

\begin{equation}
\sigma_{ab}^{\rm M2} =\left(0.2820\sigma_{aa}^6+0.4732\sigma_{aa}^3\sigma_{bb}^3+0.2448\sigma_{bb}^6\right)^{\frac{1}{6}}\,,
\label{equ_M2_sig}
\end{equation}

and

\begin{equation}
\epsilon_{ab}^{\rm M2} = \frac{0.03995\epsilon_{aa}+0.9564698845 \sqrt{\epsilon_{aa}\epsilon_{bb}}+0.00355\epsilon_{bb}} {0.2820\sigma_{aa}^6+0.4732\sigma_{aa}^3\sigma_{bb}^3+0.2448\sigma_{bb}^6} \sigma_{aa}^3\sigma_{bb}^3 \,.
\label{equ_M2_eps}
\end{equation}

In Equation (\ref{equ_M2_sig}), the lower valued LJ parameter has to be taken for site index $a$  \cite{Al-Matar2004}, since unequal weights are used. Table \ref{tab_loa} summarizes the eleven studied combining rules.

\section{Case Study \label{cs}}

Before studying the performance of the different combining rules, it is useful to get a picture of the influence of $\sigma_{ab}$ and $\epsilon_{ab}$ on the vapor-liquid equilibria of mixtures. Vapor pressure, bubble density and dew point compositon at specified temperature and bubble point composition were investigated regarding their sensitivity on $\sigma_{ab}$ and $\epsilon_{ab}$. Instead of using the absolute values of $\sigma_{ab}$ and $\epsilon_{ab}$, the unlike LJ parameters are discussed here in terms of the deviations $\eta$ and $\xi$ from the Lorentz-Berthelot combining rule 

\begin{equation}
\sigma_{ab}%^{\rm adj}
=\eta\frac{\sigma_{aa}+\sigma_{bb}}{2} \label{equ_MLB_sig}\,,
\end{equation}

and 

\begin{equation}
\epsilon_{ab}%^{\rm adj}
=\xi\sqrt{\epsilon_{aa}\epsilon_{bb}}\,. \label{equ_MLB_eps}
\end{equation}

The influence of the unlike LJ parameters on the vapor-liquid equilibria of mixtures was investigated in a case study for which two different binary mixtures ($\rm N_2$+$\rm C_3H_6$ and CO+$\rm C_2H_6$) were chosen here. Three of the regarded components were described by 2CLJQ models, only for CO a 2CLJD model was used. For CO+$\rm C_2H_6$, the case study was performed at 223~K and $x_{\rm CO}=0.198$ mol/mol, where experimental vapor pressure and dew point carbon monoxide mole fraction are 5.614~MPa and 0.8065~mol/mol \cite{Trust1971}. To eliminate the influence of experimental scatter, the pressure -- composition data were smoothed \cite{Robinson1985}. The reference bubble density of 16.499~mol/l was determined using DDMIX \cite{Ely1989} which was designed to yield reliable mixture densities.

The experimental data for $\rm N_2$+$\rm C_3H_6$ at 290~K and $x_{\rm N2}=0.1146$~mol/mol, where the case study was performed, are for the vapor pressure and dew point nitrogen mole fraction 6.783~MPa and 0.7497~mol/mol \cite{Grauso1977}. Again smoothed values were taken here \cite{Robinson1985}. The bubble density of 12.4094~mol/l was estimated by the Rackett model \cite{Rackett1970,Spencer1973} together with the mixing rule of Chueh and Prausnitz \cite{Chueh1967}.

\begin{figure}[ht]
\centering
\includegraphics[width=\textwidth]{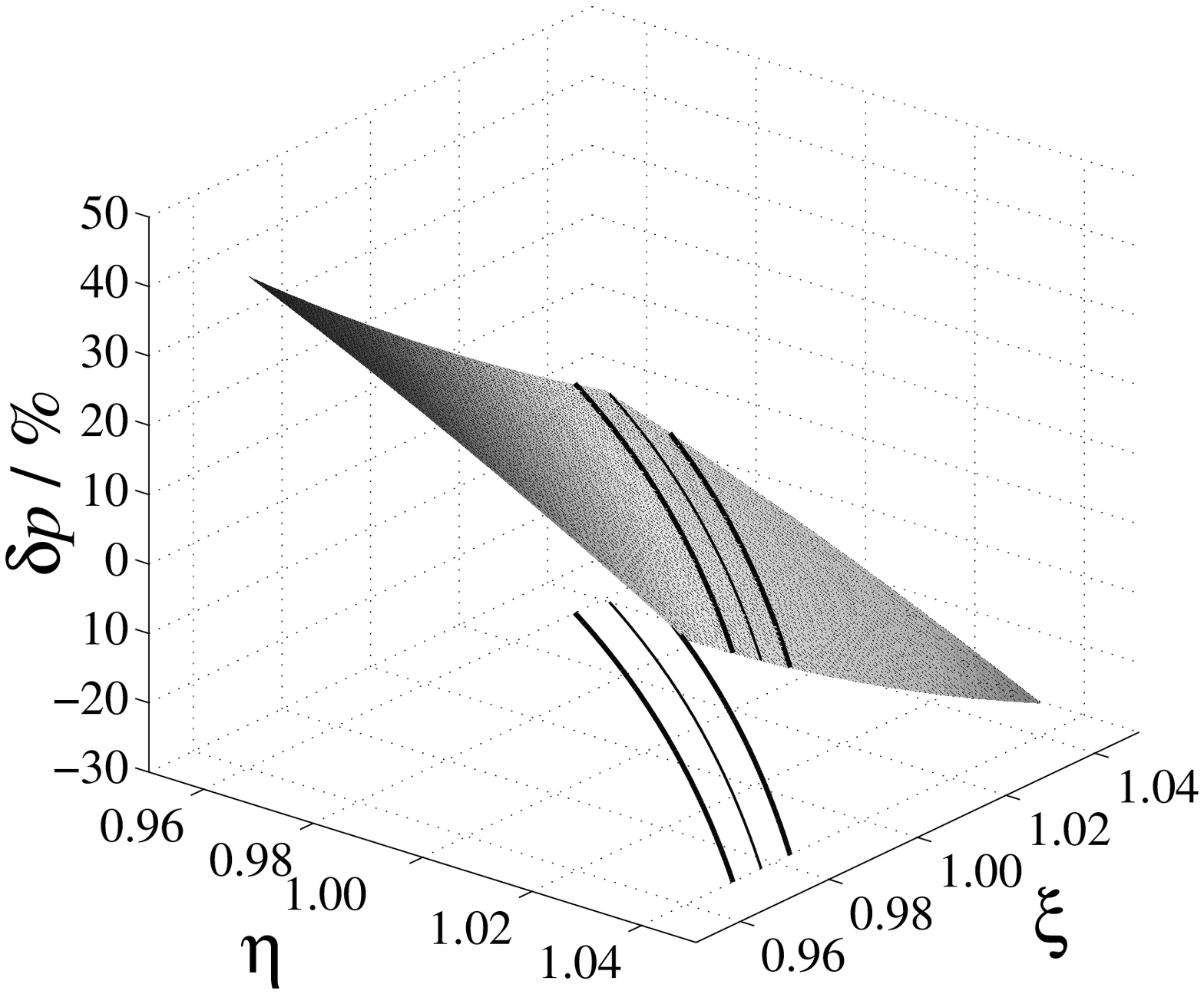}
\bigskip
\bigskip
\bigskip
\bigskip
\caption[Vapor pressure of CO+$\rm C_2H_6$ at 223~K and bubble point carbon monoxide mole fraction of 0.1980 mol/mol: deviations of the simulation results to the experimental vapor pressure of 5.614~MPa. The surface is a fit to 25 simulation results with varying $\eta$ and $\xi$. Solid slim line: $\eta$ and $\xi$ combinations yielding the experimental value. Solid bold lines: $\eta$ and $\xi$ combinations deviating by $\pm$3 \% from the experiment. Lines in the $\eta$--$\xi$ plane: projections of the lines from the surface.]{Schnabel et al.\label{fig_druck3d}}
\end{figure}

\begin{figure}[ht]
\centering
\includegraphics[width=\textwidth]{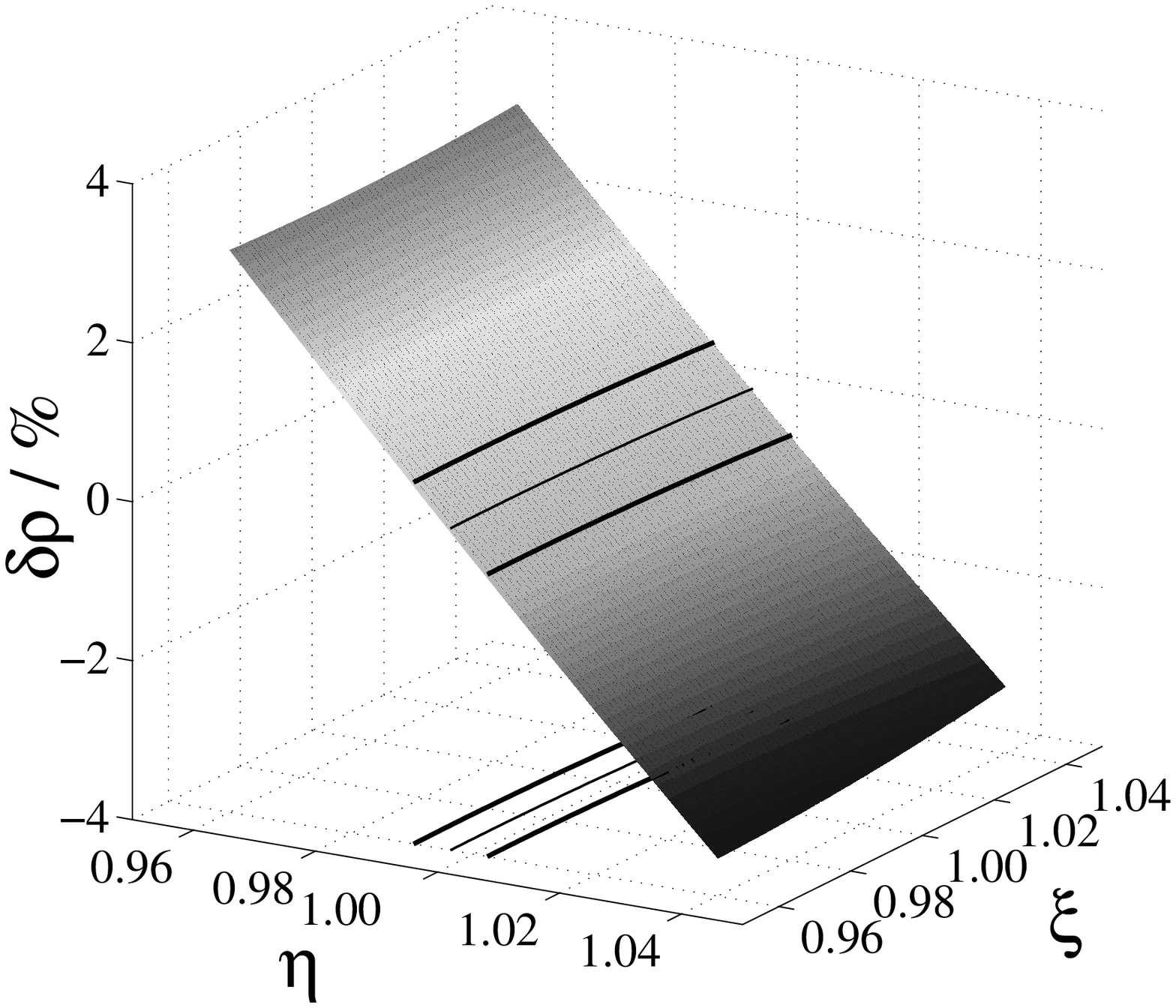}
\bigskip
\bigskip
\bigskip
\bigskip
\caption[Bubble density of CO+$\rm C_2H_6$ at 223~K and bubble point carbon monoxide mole fraction of 0.1980 mol/mol: deviations simulation results to the experimental bubble density of 16.499~mol/l. The surface is a fit to 25 simulation results with varying $\eta$ and $\xi$. Solid slim line: $\eta$ and $\xi$ combinations yielding the experimental value. Solid bold lines: $\eta$ and $\xi$ combinations deviating by $\pm$0.5 \% from the experiment. Lines in the $\eta$--$\xi$ plane: projections of the lines from the surface.]{Schnabel et al.\label{fig_dichte3d}}
\end{figure}

\begin{figure}[ht]
\centering
\includegraphics[width=\textwidth]{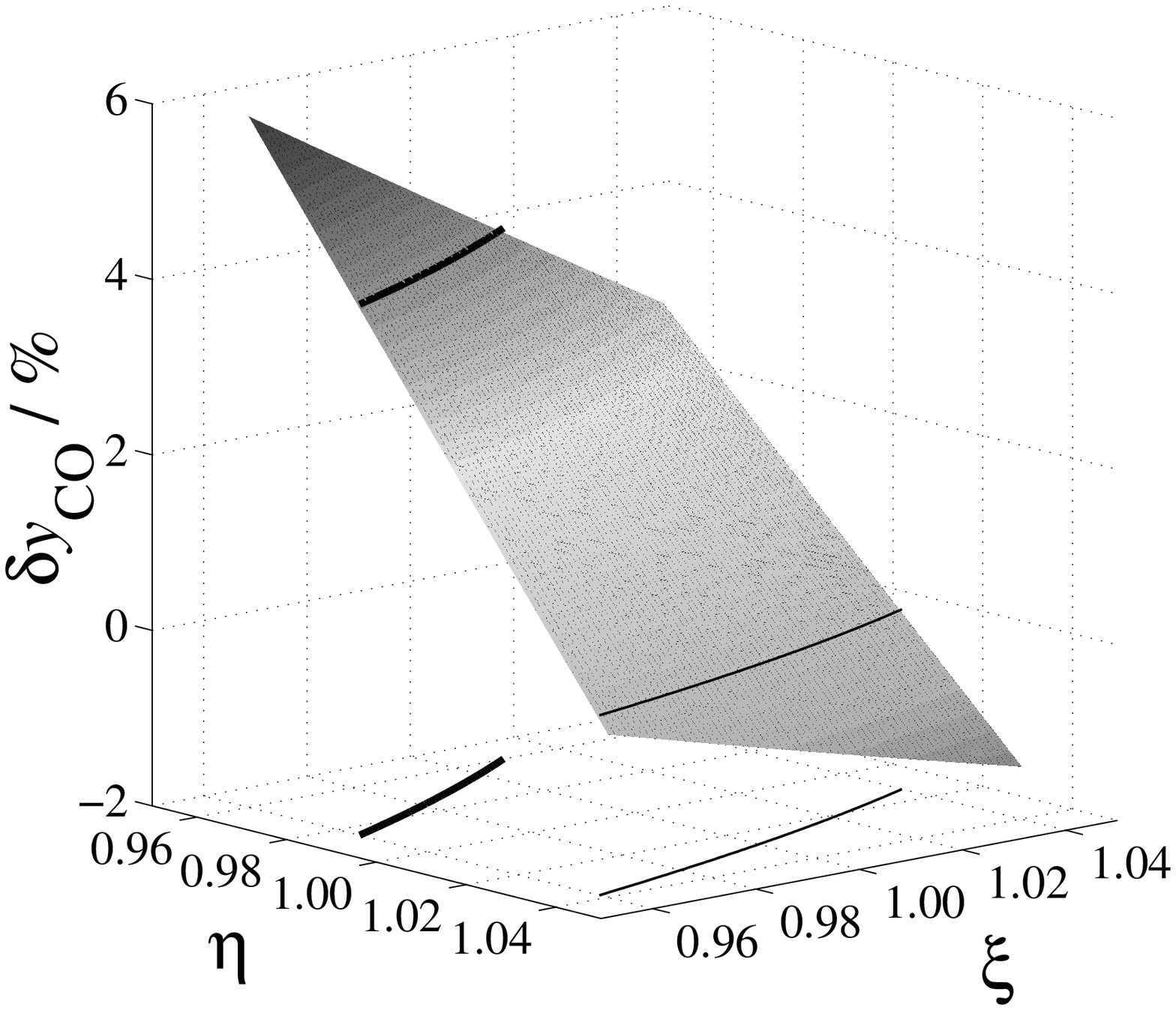}
\bigskip
\bigskip
\bigskip
\bigskip
\caption[Dew point carbon monoxide mole fraction of CO+$\rm C_2H_6$ at 223~K and bubble point carbon monoxide mole fraction of 0.1980 mol/mol: deviations of simulation results to the experimental dew point carbon monoxide composition of 0.8065 mol/mol. The surface is a fit to 25 simulation results with varying $\eta$ and $\xi$. Solid slim line: $\eta$ and $\xi$ combinations yielding the experimental value. Solid bold line: $\eta$ and $\xi$ combinations deviating by +4 \% from the experiment. Lines in the $\eta$--$\xi$ plane: projections of the lines from the surface.]{Schnabel et al.\label{fig_konz3d}}
\end{figure}

Vapor-liquid equilibria for CO+$\rm C_2H_6$ were simulated for every combination of \\
$\eta\in\left[0.96,\,0.98,\,1,\,1.02,\,1.04\right]$ and $\xi\in\left[0.96,\,0.98,\,1,\,1.02,\,1.04\right]$, i.e.~overall 25 combinations, using the Grand Equilibrium method \cite{Vrabec2002}. Simulation results of vapor pressure, bubble density and dew point composition are therefore functions of $\eta$ and $\xi$. Their deviations to the
experimental values are illustrated as three-dimensional plots in Figures \ref{fig_druck3d} to \ref{fig_konz3d} where the surfaces were derived from the 25 simulation points. All combinations of $\eta$ and $\xi$ which yield the experimental value are plotted as a solid slim line in Figures \ref{fig_druck3d} to \ref{fig_konz3d} and as a projection onto the $\eta$--$\xi$ plane. For the further discussion experimental and simulation uncertainties are needed, where the following estimates were used: 3 \% for the vapor pressure, 0.5 \% for the bubble density and 4 \% (about 0.03 mol/mol) for the dew point composition. These limits are also included in Figures \ref{fig_druck3d} to \ref{fig_konz3d} as solid bold lines. 

The vapor pressure significantly dependens on both $\eta$ and $\xi$, cf. Figure \ref{fig_druck3d}. It varies by about 35 \% for $\eta$ (at constant $\xi$) and 45 \% for $\xi$ (at contant $\eta$) in the investigated parameter range. Due to that high sensitivity, the range where the simulated and experimental vapor pressure agree within 3 \% is rather small.  

In contrast, the bubble density is practially independent of the parameter $\xi$ as can bee seen from Figure \ref{fig_dichte3d}, but it varies by about 6 \% in the investigated range of $\eta$. The experimental value can be obtained by choosing $\eta$ close to unity in combination with any value of $\xi$ in the regarded range.
 
The dew point carbon monoxide mole fraction depends both on $\eta$ and $\xi$ as shown in Figure \ref{fig_konz3d}. But it varies only by about 6 \% for $\eta$ (at constant $\xi$) and 2 \% for $\xi$ (at contant $\eta$) in the investigated parameter range.

\begin{figure}[ht]
\centering
\includegraphics[width=\textwidth]{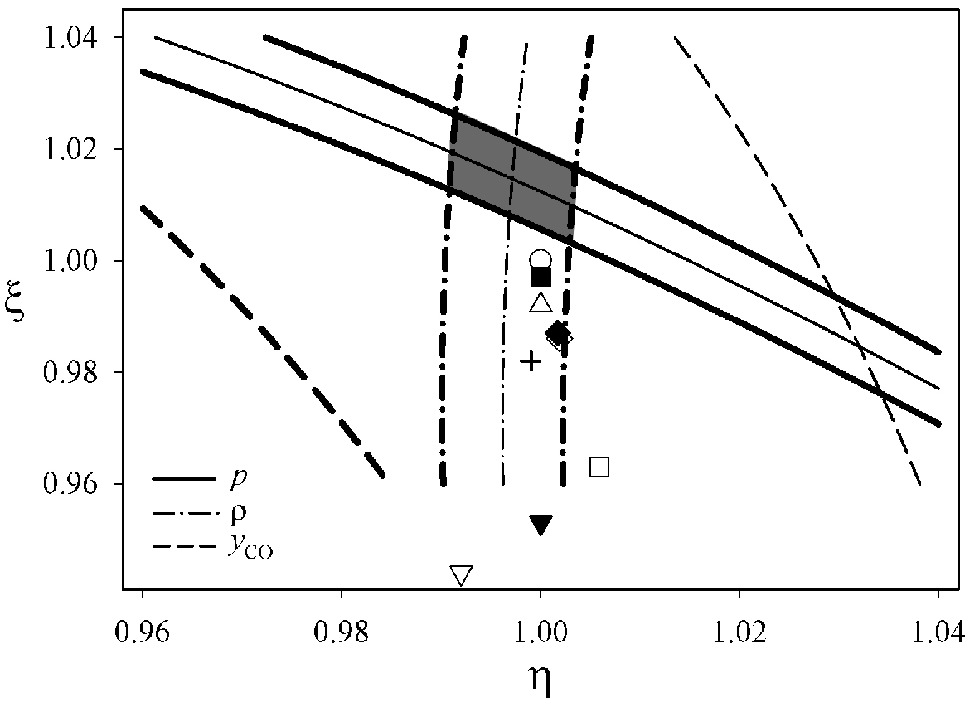}
\bigskip
\bigskip
\bigskip
\bigskip
\caption[Superposition of the projections from Figures \ref{fig_druck3d} to \ref{fig_konz3d} (CO+$\rm C_2H_6$). The thin lines indicate the combinations of $\eta$ and $\xi$ where simulation and experiment coincide for the different properties. The thick lines represent the assumed uncertainties of $\delta p=\pm3$ \%,  $\delta\rho=\pm0.5$ \% and $\delta y_{\rm{}_{CO}}=+4$ \%. The predictions of the combinining rules are given by: {\Large $\circ$} LB, $\blacksquare$ K, {$\vartriangle$} HMC, $\triangledown$ H, $\blacktriangledown$ S, $\square$ SK, {\Large $\diamond$} WH, $\blacklozenge$ M1, + M2. The remaining two combining rules are not within the scale. Shaded: target area of $\eta$ and $\xi$ combinations yielding experimental values within their uncertainties.]{Schnabel et al.\label{fig_loe1}}
\end{figure}

Summarizing Figures \ref{fig_druck3d} to \ref{fig_konz3d} in Figure \ref{fig_loe1}, it can be seen that there is one combination of $\eta$ and $\xi$ where both vapor pressure and bubble density of molecular mixture model and experiment coincide exactly ($\eta=0.9972$ and $\xi=1.0145$). With the uncertainties for the bubble density and the vapor pressure, a target area is defined in which the combination of $\eta$ and $\xi$ yields an accurate description of these both properties, cf.~the shaded area in Figure \ref{fig_loe1}. The experimental dew point composition within its (large) uncertainty is also met in this target area.

Figure \ref{fig_loe1} also allows an easy assessment of the combining rules introduced in Section \ref{coru}. The LB combining rule is the closest to the target area, but outside of it. All combining rules shown, except SK, describe the bubble density and dew point composition within their assigned uncertainty. However, the experimental vapor pressure is predicted by none of them. The LB combining rule ($\eta=1$ and $\xi=1$) yields deviations from the experimental vapor pressure, bubble density and dew point carbon monoxide mole fraction of $+5.6$ \%, $-0.3$ \% and $+1.8$ \%, respectively. The SK combining rule gives the most remote combination ($\eta=1.0059$ and $\xi=0.9630$) with deviations of $+21.7$ \%, $-0.8$ \% and $+2.3$ \%, respectively.

\begin{figure}[ht]
\centering
\includegraphics[width=\textwidth]{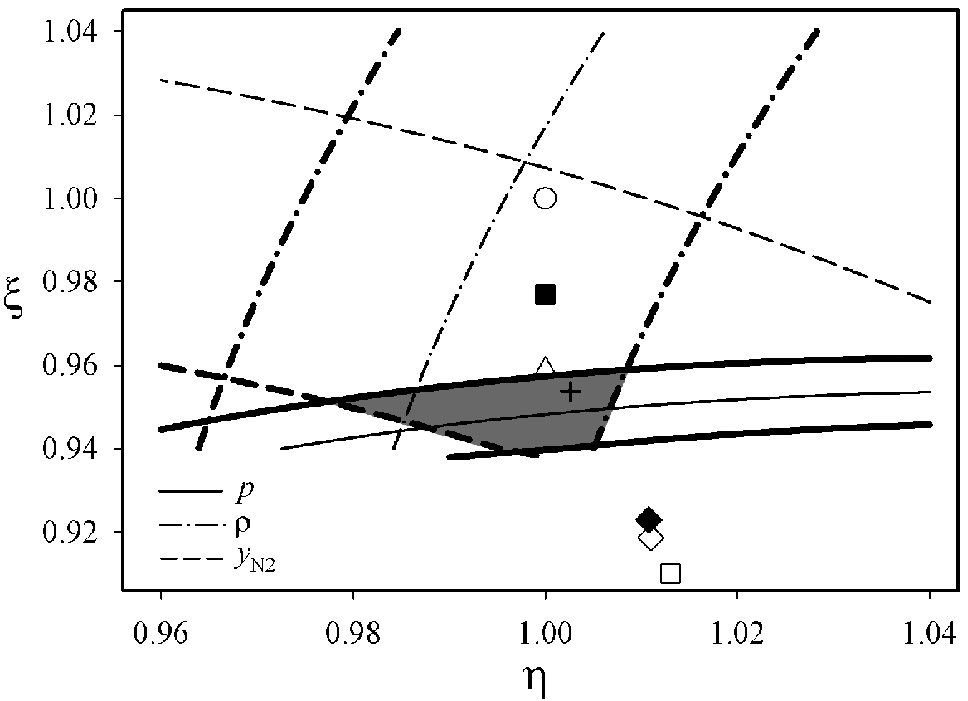}
\bigskip
\bigskip
\bigskip
\bigskip
\caption[Plot analogous to Figure \ref{fig_loe1} for $\rm N_2$+$\rm C_3H_6$ at 290~K and bubble point nitrogen mole fraction of 0.1146 mol/mol. The thin lines indicate the combinations of $\eta$ and $\xi$ where simulation and experiment coincide for the different properties. The thick lines represent the assumed uncertainties of $\delta p$=$\pm3$ \%, $\delta\rho=\pm1$ \% and $\delta y_{\rm{}_{N2}}=+4$ \%. The predictions of the combinining rules are given by: {\Large $\circ$} LB, $\blacksquare$ K, {$\vartriangle$} HMC, $\square$ SK, {\Large $\diamond$} WH, $\blacklozenge$ M1, + M2. The remaining four combining rules are not within the scale. Shaded: target area of $\eta$ and $\xi$ combinations yielding experimental values within their uncertainties.]{Schnabel et al.\label{fig_loe2}}
\end{figure}

An analogous systematic study for $\rm N_2$+$\rm C_3H_6$ includes 30 vapor-liquid equilibria simulation results for $\eta\in\left[0.96,\,0.98,\,1,\,1.02,\,1.04\right]$ and $\xi\in\left[0.94,0.96,\,0.98,\,1,\,1.02,\,1.04\right]$ at 290~K and $x_{\rm N2}=0.1146$ mol/mol. Since the experimental bubble density was estimated with the Racket model, a higher uncertainty of 1 \% was assigned to it. Figure \ref{fig_loe2} shows the comparison to experimental data where the target area is obtained analogously to the study of CO+$\rm C_2H_6$. Only the M2 combining rule is within this area, whereas HMC is very close to it. For M2 ($\eta=1.0026$, $\xi=0.9538$) deviations from the experimental vapor pressure, bubble density and dew point nitrogen mole fraction of $-1.7$ \%, $-0.8$ \% and $+3.0$ \% were found, respectively. SK ($\eta=1.013$, $\xi=0.9101$) yields again the most remote results with deviations of $+15.5$ \%, $-1.6$ \% and $+4.8$ \%, respectively.

From both case studies, the conclusion is drawn that the vapor pressure, which significantly depends on both unlike LJ parameters, is the most difficult property to be predicted by combining rules. The bubble density is mainly determined by the unlike LJ size parameter and values close to the arithmetic mean ($\eta=1$) describe the experimental bubble densities well (see also \cite{Vrabec1995,Vrabec1996}). This is achieved by most of the combining rules. The dew point composition depends on both $\eta$ and $\xi$ like the vapor pressure, however, its sensitivity is considerably weaker compared to the vapor pressure. Due to these findings, it can be recommended to use the Lorentz rule for the size parameter since it predicts bubble densities well and to adjust the unlike energy parameter to the vapor pressure.

\section{Comprehensive Study \label{os}}

In previous work \cite{Stoll2003,Vrabec2004,Stoll2004}, molecular models for 44 binary mixtures were developed. For describing the mixture, the Lorentz rule was used for $\sigma_{ab}$, while $\epsilon_{ab}$ was determined using the adjustment parameter $\xi$ as defined by Equation (\ref{equ_MLB_eps}). The state independent parameter $\xi$ was optimized by a fit to one experimental vapor pressure data point of the mixture. That procedure chosen in \cite{Stoll2003,Vrabec2004,Stoll2004} is impressively supported by the results of the case study presented above.

The results from \cite{Stoll2003,Vrabec2004,Stoll2004} can be directly used to assess combining rules. For combining rules which employ the Lorentz rule for $\sigma_{ab}$ this is completely straight-forward as the adjusted $\xi$ parameter is known and only needs to be compared to the results for the different combining rules. But also combining rules that do not employ the Lorentz rule can be included in the assessment as can be seen from the results of the case study and explained below in more detail.

The pure components of the 44 binary mixtures are of three different molecular model types: spherical and non-polar (1CLJ), anisotropic and dipolar (2CLJD) and anisotropic and quadrupolar (2CLJQ). Hence, six mixture types can be distinguished as given by Table \ref{tab_mix_type}. The adjusted unlike LJ energy parameters of those 44 mixture models deviate by up to $\pm 10$ \% from the Berthelot rule, cf.~Table \ref{tab_uLJeps} or see \cite{Stoll2003,Vrabec2004,Stoll2004}.

Seven of the eleven combining rules do not use the Lorentz rule for the unlike LJ size parameter. For an assessment, it has to be investigated by how much they deviate from the Lorentz rule. This was done here with the root mean squares (RMS)   

\begin{equation}
{\rm RMS}^{\rm x}_{\sigma}=\sqrt{\frac{1}{N}\sum_{i=1}^{N}\left(\eta^{\rm x}-1\right)^2}\,,
\label{equ_rms_sigma}
\end{equation}

\noindent 
where $N$ is the number of mixture models to which the particular combining rules (x = H, S, SK, HHG, WH, M1, M2) were applied. In Table \ref{tab_rms_sig}, ${\rm RMS}^{\rm x}_{\sigma}$ are given distinguishing between the mixture types 1 to 6. It additionally incorporates data taking all 44 binary mixture models into account. 

Table \ref{tab_rms_sig} shows that the S combining rule yields the same unlike LJ size parameters as the Lorentz rule for all mixture types. Also M2 and SK are very close, deviating by approximately 0.5 \% only. The remainder, H, HHG, WH and M1, yield all mean deviations in a narrow band at 1.4 \% and below. For mixture type 6, all seven combining rules yield results very close to the arithmetic mean due to fact that all components have very similar parameters. Figures \ref{fig_loe1} and \ref{fig_loe2} give an impression of the influence of a variation of $\eta$ of 1 \% on the mixture vapor-liquid equilibria. For the pressure and the dew point composition that variation will typically result in changes that do not exceed the experimental uncertainty. For the bubble density the changes are more significant but still the changes do not exceed 1 \% in the bubble density. Thus, it can be concluded that there is only a small influence of the combining rules via $\sigma_{ab}$ on the vapor-liquid equilibrium. Therefore, in the following only results of the combining rules for $\xi$ are discussed.

\begin{figure}[ht]
\centering
\includegraphics[width=\textwidth]{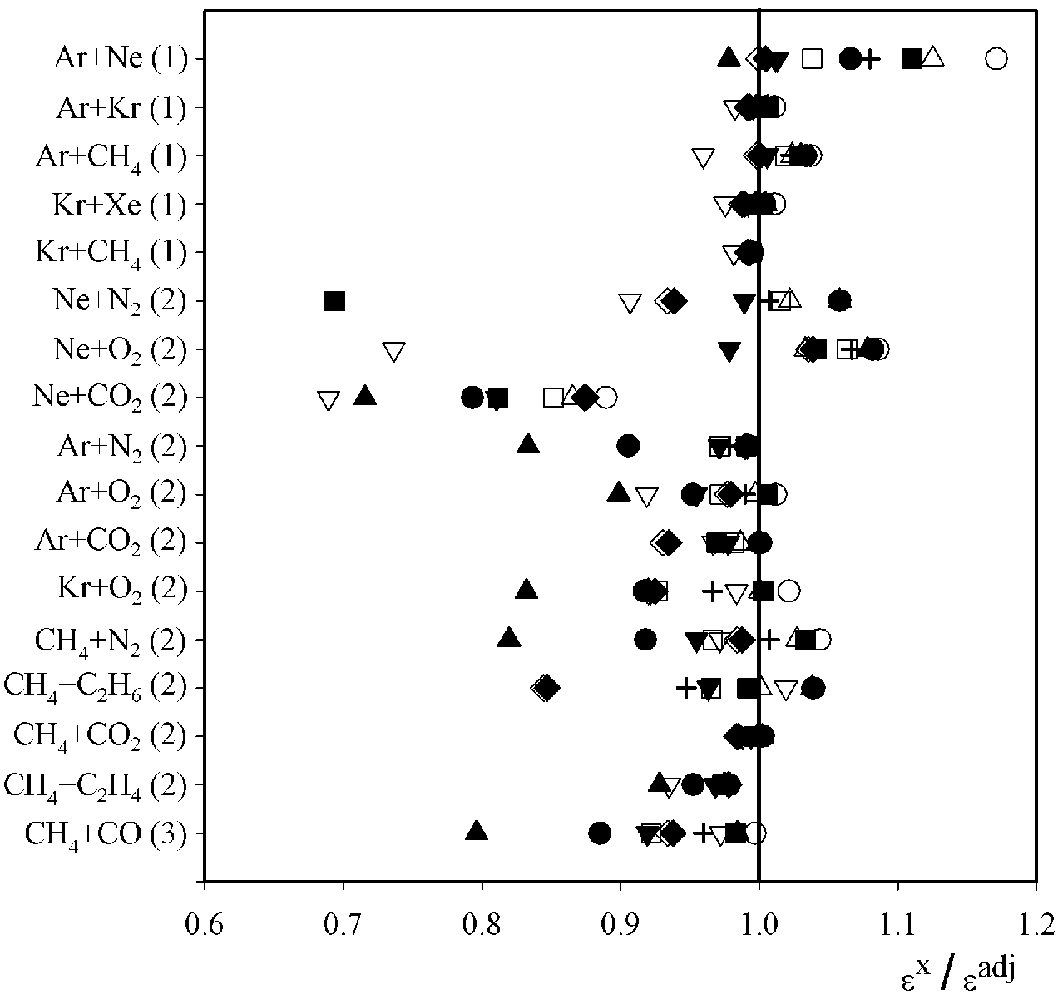}
\bigskip
\bigskip
\bigskip
\bigskip
\caption[Predicted unlike LJ energy parameters $\epsilon_{ab}^{\rm x}$ from eleven combining rules compared to adjusted values for the mixture types 1CLJ+1CLJ (1), 1CLJ+2CLJQ (2) and 1CLJ+2CLJD (3): {\Large $\circ$} LB, $\blacksquare$ K, {$\vartriangle$} HMC, $\blacktriangle$ FH, $\triangledown$ H, $\blacktriangledown$ S, $\square$ SK, {\Large $\bullet$} HHG, {\Large $\diamond$} WH, $\blacklozenge$ M1, + M2.]{Schnabel et al.\label{fig_Tpy1_2_3}}
\end{figure}

Unlike LJ energy parameters predicted by the eleven combining rules are compared
to the adjusted values in Figure \ref{fig_Tpy1_2_3}. The adequacy of the investigated combining rules is indicated by the distance from the reference line (unity), which represents the adjusted LJ energy parameter. Generally, too low dispersion energies are predicted, which is particlularly visible for mixture types 2 to 5. Note that it is implicitly assumed in Figure \ref{fig_Tpy1_2_3} that the adjusted parameter for $\epsilon_{ab}$ determined in \cite{Stoll2003,Vrabec2004,Stoll2004} using the Lorentz rule for $\sigma_{ab}$ also holds for the combining rules which apply other approaches for determining $\sigma_{ab}$.

To assess the predictive power of the eleven combining rules the deviations of the unlike LJ energy parameters of the particular combining rules from the adjusted values were summarized by the root mean square

\begin{equation}
{\rm RMS}^{\rm x}_{\epsilon}=\sqrt{\frac{1}{N}\sum_{i=1}^{N}\left(\xi^{\rm x}-\xi^{\rm adj}\right)^2}\,.
\label{equ_rms}
\end{equation}

The deviations for all eleven combining rules, distinguishing between mixture types, are given in Table \ref{tab_rms_eps}. For mixture type 1, for which most of the eleven combining rules were tested, they are better than the LB combining rule. Many of them were fitted to noble gas mixture data and they indeed predict good unlike LJ energy paramters for such mixtures. The reason is that all combining rules, except LB and M2, yield smaller values than the geometric mean. Such a lower unlike dispersion energy is supported by experimental noble gas mixture data \cite{Tang1986,Halgren1992,Kestin1984,Bzowski1990}. However, for all other mixture types 2 to 5, the combining rules can predict better or worse than the simple LB combining rule. That all combining rules predict similar for mixture type 6 is due to the very similar values of both LJ parameters of the pure components. All in all, the simple LB combining rule has a ${\rm RMS}^{\rm x}_{\epsilon}$ of 5 \% which is not significantly worse than those of the best rule (HMC) with 3.9 \%, where a variation of $\xi$ by 1 \% influences the vapor pressure by about 3 \%. 

Apart from noble gas mixtures, it can be concluded that none of the inverstigated combining rules has a significant advantage over LB, which yields good bubble densities and vapor compositions, but vapor pressures that deviate from the experiment by an estimated 15 \% on average. If highly accurate results for the vapor pressure are needed, an adjustment of the unlike LJ energy parameter to at least one experimental binary vapor pressure data point is necessary in most cases.

\section{Conclusion}

The dependence of vapor-liquid equilibrium properties on unlike LJ parameters was studied systematically for the mixtures CO+$\rm C_2H_6$ and $\rm N_2$+$\rm C_3H_6$. This case study shows that the mixture bubble density is accurately obtained using the arithmetic mean of the like LJ size parameters as proposed by the Lorentz combining rule. The bubble density is insensitive to variations of the unlike LJ energy parameter. The vapor pressure is found to be dependent on both unlike LJ parameters. The same is found for vapor phase composition, but with a considerably lower sensitivity. Therefore, it can be recommended to use the Lorentz rule for the unlike LJ size parameter and to adjust the unlike LJ energy parameter to the vapor pressure. Recent investigations on mixtures consisting of more complex molecules, such as ethanol \cite{Schnabel2005,Schnabel2006}, confirm that this procedure holds also in such cases.

Eleven combining rules for unlike LJ parameters, some of which use also other properties than the pure component LJ parameters like ionization potentials or
polarizabilities, were investigated regarding their predictive power for vapor-liquid equilibria. They are compared to 44 adjusted mixture models from previous work \cite{Stoll2003,Vrabec2004,Stoll2004}, where the unlike LJ energy parameters were optimized to yield experimental vapor pressure. It is found that the unlike LJ size parameters from the combining rules differ only very litte from the Lorentz rule. Many combining rules investigated here were fitted to noble gas mixture data and they indeed predict good unlike LJ energy paramters for such mixtures. However, none of the investigated combining rules has a significant advantage over LB for other mixture types.

% Acknowledgment
\section{Acknowledgment}

The authors gratefully acknowledge financial support by Deutsche Forschungsgemeinschaft, Schwerpunktprogramm 1155.

\clearpage

% References

%\bibliography{/usr/ITT/schnabel/Literatur/combining_rule}

\begin{thebibliography}{10}

\bibitem{Pena1980}
M. Diaz Pe\~na, C. Pando, J.A.R. Renuncio,
%\newblock Combination Rules for Two-Body van der Waals Coefficients.
\newblock  J. Chem. Phys. 72 (1980) 5269--5275.

\bibitem{Tang1986}
K.T. Tang, J.P. Toennies,
%\newblock New Combining Rules for Well Parameters and Shapes of the van der Waals Potential of Mixed Rare Gas Systems.
\newblock  Zeitschrift f{\"u}r Physik D 1 (1986) 91--101.

\bibitem{Bzowski1988}
J. Bzowski, E.A. Mason, J. Kestin,
%\newblock On Combination Rules for Molecular Van der Waals Potential-Well Parameters.
\newblock  Int. J. Thermophys. 9 (1988) 131--143.

\bibitem{Mason1964}
E.A. Mason, M. Islam, S. Weissmann,
%\newblock Thermal Diffusion and Diffusion in Hydrogen-Krypton Mixtures.
\newblock  Phys. Fluids 7 (1964) 1011--1022.

\bibitem{Good1971}
R.J. Good, C.J. Hope,
%\newblock Test of Combining Rules for Intermolecular Distances. Potenial function constants from second virial coefficients.
\newblock  J. Chem. Phys. 55 (1964) 111--116.

\bibitem{Pena1982}
M. Diaz Pe\~na, C. Pando, J.A.R. Renuncio,
%\newblock Combination Rules for Intermolecular Potential Parameters. I. Rules Based on Approximations for the Long-Range Dispersion Energy.
\newblock  J. Chem. Phys. 76 (1964) 325--332.

\bibitem{Pena1982a}
M. Diaz Pe\~na, C. Pando, J.A.R. Renuncio,
%\newblock Combination Rules for Intermolecular Potential Parameters. II. Rules Based on Approximations for the Long-Range Dispersion Energy and an Atomic Distortion Model for the Repulsive Interactions.
\newblock  J. Chem. Phys. 76 (1964) 333--339.

\bibitem{Kohler1982}
F. Kohler, J. Fischer, E. Wilhelm,
%\newblock Intermolecular Force Parameters for Unlike Pairs.
\newblock  J. Mol. Struct. 84 (1982) 245--250.

\bibitem{Moeller1992}
D. M{\"o}ller, J. \'Oprzynski, A. M{\"u}ller, J. Fischer,
%\newblock Prediction of Thermodynamic Properties of Fluid Mixtures by   Molecular-Dynamics Simulations: Methane-Ethane.
\newblock  Mol. Phys. 75 (1992) 363--378.

\bibitem{Vrabec1995}
J. Vrabec, J. Fischer, 
%\newblock Vapour Liquid Equilibria of Mixtures from the $NpT$ plus Test Particle Method.
\newblock  Mol. Phys. 85 (1995) 781--792.

\bibitem{Vrabec1996}
J. Vrabec, J. Fischer.
%\newblock Vapor-Liquid Equilibria of Binary Mixtures Containing Methane, Ethane, and Carbon Dioxide from Molecular Simulation. 
\newblock  Int. J. Thermophys. 17 (1996) 889--908.

\bibitem{Kronome2000}
G. Kronome, I. Szalai, M. Wendland, J. Fischer,
%\newblock Extension of the $NpT$ + Test Particle Method for the Calculation of Phase Equilibria of Nitrogen + Ethane.
\newblock  J. Mol. Liq. 85 (2000) 237--247.

\bibitem{Stoll2003}
J. Stoll, J. Vrabec, H. Hasse,
%\newblock Vapor-liquid Equilibria of Mixtures Containing Nitrogen, Oxygen, Carbon Dioxide, and Ethane.
\newblock  AIChE J. 49 (2000) 2187--2198.

\bibitem{Vrabec2004}
J. Vrabec, J. Stoll, H. Hasse,
%\newblock Molecular Models of Unlike Interactions in Mixtures.
\newblock  Mol. Simul. 31 (2005) 215--221.

\bibitem{Stoll2004}
J. Stoll,
\newblock Molecular Models for the Prediction of Thermophysical Properties of Pure Fluids and Mixtures,
\newblock  Fortschritt-Berichte VDI, Reihe 3, 836, VDI Verlag, D\"usseldorf 2004.

\bibitem{Vrabec2001}
J. Vrabec, J. Stoll, H. Hasse,
%\newblock A Set of Molecular Models for Symmetric Quadrupolar Fluids.
\newblock  J. Phys. Chem. B 105 (2001) 12126--12133.

\bibitem{Stoll2003a}
J. Stoll, J. Vrabec, H. Hasse,
%\newblock A Set of Molecular Models for Carbon Monoxide and Halogenated Hydrocarbons.
\newblock  J. Chem. Phys. 119 (2003) 11396--11407.

\bibitem{Vrabec2004a}
J. Vrabec, G.K. Kedia, H. Hasse,
%\newblock Prediction of Joule-Thomson Inversion Curves for Pure Fluids and One Mixture by Molecular Simulation. 
\newblock  Cryogenics 45 (2005) 253--258.

\bibitem{Fernandez2004}
G.A. Fern\'andez, J. Vrabec and H. Hasse, 
%\newblock Self diffusion and binary Maxwell-Stefan diffusion in simple fluids with the Green-Kubo method. 
\newblock  Int. J. Thermophys. 25 (2004) 175--186.

\bibitem{Fernandez2004a}
G.A. Fern\'andez, J. Vrabec, H. Hasse,
%\newblock A molecular simulation study of shear and bulk viscosity and thermal conductivity of simple real fluids. 
\newblock  Fluid Phase Equilib. 221 (2004) 157--163.

\bibitem{Vrabec1997a}
J. Vrabec, J. Fischer,
%\newblock Vapor-Liquid Equilibria of the Ternary Mixture $\rm CH_4$+$\rm C_2H_6$+$\rm CO_2$ from Molecular Simulation.
\newblock  AIChE J. 43 (1997) 212--217.

\bibitem{Lorentz1881}
H.A. Lorentz,
%\newblock {\"U}ber die Anwendung des Satzes vom Virial in der kinetischen Theorie der Gase.
\newblock  Annalen der Physik 12 (1881) 127--136.

\bibitem{Berthelot1898}
D. Berthelot,
%\newblock Sur le Mélange des Gaz.
\newblock  Comptes Rendus de l'Académie des Sciences Paris 126 (1889) 1703--1706.

\bibitem{Delhommelle2001}
J. Delhommelle, P. Milli\'e,
%\newblock Inadequacy of the Lorentz-Berthelot Combining Rules for Accurate Predictions of Equilibrium Properties by Molecular Simulation.
\newblock  Mol. Phys. 99 (2001) 619--625.

\bibitem{Ungerer2004}
P. Ungerer, A. Wender, G. Demoulin, E. Bourasseau, P. Mougin,
%\newblock Application of Gibbs Ensemble and $NpT$ Monte Carlo Simulation to the Development of Improved Processes for $\rm H_2S$-rich Gases.
\newblock  Mol. Simul. 30 (2004) 631--648.

\bibitem{Kohler1957}
F. Kohler,
%\newblock Zur Berechnung der Wechselwirkungsenergie zwischen ungleichen Molek{\"u}len in binären fl{\"u}ssigen Mischungen.
\newblock  Monatsh. Chem. 88 (1957) 857--877.

\bibitem{London1930}
F. London,
%\newblock \"Uber einige Eigenschaften und Anwendungen der Molekularkr\"afte.
\newblock  Z. Phys. Chem. (Leipzig, Ger.) 11 (1930) 222--251.

\bibitem{London1937}
F. London,
%\newblock The General Theory of Molecular Forces.
\newblock  Transactions of the Faraday Society 33 (1937) 8--26.

\bibitem{Laidig1990}
K.E. Laidig, R.F.W. Bader,
%\newblock Properties of Atoms in Molecules: Atomic polarizabilities.
\newblock  J. Chem. Phys. 93 (1990) 7213--7224.

\bibitem{Lide1995-1996}
R.D. Lide,
\newblock  CRC Handbook of Chemistry and Physics, 76th ed.,
\newblock CRC Press Inc., New York, 1995-1996.

\bibitem{Wood2002}
C. D. Wood, K. Senoo, C. Martin, J. Cuellar and A.I. Cooper,
%\newblock Polymer Synthesis Using Hydrofluorocarbon Solvents. 1. Synthesis of Cross-Linked Polymers by Dispersion Polymerization in 1,1,1,2-Tetrafluoroethane.
\newblock  Macromolecules 35 (2002) 6743--6746.

\bibitem{Fermeglia2003}
M. Fermeglia, M. Ferrone, S. Pricl,
%\newblock Development of An All-Atoms Force Field from ab initio Calculations for Alternative Refrigerants.
\newblock  Fluid Phase Equilib. 210 (2003) 105--116.

\bibitem{Echt1988}
O. Echt, D. Kreisle, E. Recknagel, J.J. Saenz, R. Casero and J.M. Soler,
%\newblock Dissociation Channels of Multiply Charged van der Waals Clusters.
\newblock  Phys. Rev. A 38 (1988) 3236--3248.

\bibitem{Hudson1960}
G.H. Hudson, J.C. McCoubrey,
%\newblock Intermolecular Forces between Unlike Molecules. A More Complete Form of the Combining Rules.
\newblock  Transactions of the Faraday Society 56 (1960) 761--766.

\bibitem{Reed1955a}
T.M. Reed, III,
%\newblock The Theoretical Energies of Mixing for Fluorocarbon-Hydrocarbon Mixtures.
\newblock  J. Phys. Chem. 59 (1955) 425--428.

\bibitem{Srivastava1953}
B.N. Srivastava, M.P. Madan,
%\newblock Thermal Diffusion of Gas Mixtures and Forces between Unlike Molecules.
\newblock  Proceedings of the Physical Society of London Section A 66 (1953) 278--287.

\bibitem{Reid1965}
R.C. Reid, T.W. Leland, 
%\newblock Pseudocritical Constants.
\newblock  AIChE J. 11 (1965) 228--237.

\bibitem{Fender1962}
B.E.F. Fender and G.D. Halsey,
%\newblock Second Virial Coefficients of Argon, Krypton, and Argon-Krypton Mixtures at Low Temperatures.
\newblock  J. Chem. Phys. 36 (1962) 1881--1888.

\bibitem{Hiza1969}
M.J. Hiza, A.G. Duncan,
%\newblock Comments on ``Intermolecular Forces: Thermal Diffusion and Diffusion in He-Kr and $\rm H_2-Kr$''.
\newblock  Phys. Fluids 12 (1969) 1531--1532.

\bibitem{Hiza1970}
M.J. Hiza, A.G. Duncan,
%\newblock A Correlation for the Prediction of Interaction Energy Parameters for Mixtures of Small Molecules.
\newblock  AIChE J. 16 (1970) 733--738.

\bibitem{Hiza1978}
M.J. Hiza, R.L. Robinson,
%\newblock Comment on ``Intermolecular forces in mixtures of helium with the heavier noble gases''.
\newblock  J. Chem. Phys. 68 (1978) 4768--4769.

\bibitem{Sikora1970}
P.T. Sikora,
%\newblock Combining Rules for Spherically Symmetric Intermolecular Potentials.
\newblock  J. Phys. B 3 (1970) 1475--1482.

\bibitem{Maitland1981}
G.C. Maitland,
\newblock Intermolecular Forces,
\newblock Clarendon Press, Oxford, 1981.

\bibitem{Smith1972}
F.T. Smith,
%\newblock Atomic Distortion and the Combining Rule for Repulsive Potentials.
\newblock  Phys. Rev. A 5 (1972) 1708--1713.

\bibitem{Kong1973}
C.L. Kong,
%\newblock Combining Rules for Intermolecular Potential Parameters. II. Rules for the Lennard-Jones (12-6) Potential and the Morse Potential.
\newblock  J. Chem. Phys. 59 (1973) 2464--2467.

\bibitem{Halgren1992}
T.A. Halgren,
%\newblock Representation of van der Waals (vdW) Interactions in Molecular Mechanics Force Fields: Potential form, combination rules, and vdW parameters.
\newblock  J. Am. Chem. Soc. 114 (1992) 7827--7843.

\bibitem{Waldman1993}
M. Waldman, A.T. Hagler,
%\newblock New Combining Rules for Rare Gas van der Waals Parameters.
\newblock  J. Comput. Chem. 14 (1993) 1077--1084.

\bibitem{Al-Matar2004}
A.K. Al-Matar, D.A. Rockstraw,
%\newblock A Generating Equation For Mixing Rules and Two New Mixing Rules for Interatomic Potential Energy Parameters.
\newblock  J. Comput. Chem. 25 (2004) 660--668.

\bibitem{Trust1971}
D.B. Trust and F. Kurata,
%\newblock Vapor-Liquid and Liquid-Liquid Vapor Phase Behavior of the Carbon Monoxide-Propane and the Carbon Monoxide-Ethane Systems.
\newblock  AIChE J. 17 (1971) 415--419.

\bibitem{Robinson1985}
D.B. Robinson, D.-Y. Peng, S.Y.-K. Chung,
%\newblock The Development of the Peng-Robinson Equation and Its Application to Phase Equilibrium in a System Containing Methanol.
\newblock  Fluid Phase Equilib. 24 (1985) 25--41.

\bibitem{Ely1989}
J.F. Ely, J.W. Magee, W.M. Haynes,
\newblock NBS standard reference database 14, DDMIX, Version 9.06.
\newblock {NIST}, Boulder, Colorado 1989.

\bibitem{Grauso1977}
L. Graus{\o}, A. Fredenslund, J. Mollerup,
%\newblock Vapor-Liquid Equilibrium Data for the Systems $\rm C_2H_6+N_2$, $\rm C_2H_4+N_2$, $\rm C_3H_8+N_2$, and $\rm C_3H_6+N_2$.
\newblock  Fluid Phase Equilib. 1 (1977) 13--26.

\bibitem{Rackett1970}
H.G. Rackett,
%\newblock Equation of State for Saturated Liquids.
\newblock  Journal of Chemical and Engineering Data 15 (1970) 514--517.

\bibitem{Spencer1973}
C.F. Spencer, R.P. Danner,
%\newblock Prediction of Bubble-Point Density of Mixtures.
\newblock  Journal of Chemical and Engineering Data 18 (1973) 230--234.

\bibitem{Chueh1967}
P.L. Chueh, J.M. Prausnitz,
%\newblock Vapor-Liquid Equilibria at High Pressures: Calculation of Partial Molar Volumes in Nonpolar Liquid Mixtures.
\newblock AIChE J. 13 (1967) 1099--1107.

\bibitem{Vrabec2002}
J. Vrabec, H. Hasse,
%\newblock Grand Equilibrium: vapour-liquid equilibria by a new molecular simulation method.
\newblock  Mol. Phys. 100 (2002) 3375--3383.

\bibitem{Kestin1984}
J. Kestin, K. Knierim, E.A. Mason, B. Najafi, S.T. Ro, M. Waldman,
%\newblock Equilibrium and Transport Properties of the Noble Gases and Their Mixtures at Low Density.
\newblock  J. Phys. Chem. Ref. Data 13 (1984) 229--303.

\bibitem{Bzowski1990}
J. Bzowski, J. Kestin, E.A. Mason, F.J. Uribe,
%\newblock Equilibrium and Transport Properties of Gas mixtures at Low Density: Eleven polyatomic gases and five noble gases.
\newblock  J. Phys. Chem. Ref. Data 19 (1990) 1179--1232.

\bibitem{Schnabel2005}
T. Schnabel, J. Vrabec, H. Hasse,
%\newblock Henry's Law Constants of Methane, Nitrogen, Oxygen and Carbon Dioxide in Ethanol from 273 to 498 K: Prediction from molecular simulations.
\newblock Fluid Phase Equilib. 233 (2005) 134--143.

\bibitem{Schnabel2006}
T. Schnabel, J. Vrabec, H. Hasse,
%\newblock Erratum to Henry's Law Constants of Methane, Nitrogen, Oxygen and Carbon Dioxide in Ethanol from 273 to 498 K: Prediction from molecular simulations [Fluid Phase Equilib. 233 (2005) 134--143].
\newblock Fluid Phase Equilib. 239 (2006) 125--126.

\end{thebibliography}
%\bibliographystyle{unsrt}

% Tables

\clearpage

\begin{table}[ht]
\noindent
\caption[]{Eleven selected combining rules together with the acronyms used here and corresponding references. In the last column, it is indicated whether the unlike LJ size parameter is given by the arithmetic mean, i.e.~the Lorentz rule. \label{tab_loa}}
\bigskip
\begin{center}
\begin{tabular}{|l|l|l|c|} \hline

Name              & Acronym & Reference                        &  Lorentz  \\\hline
Lorentz-Berthelot & LB    & \cite{Lorentz1881,Berthelot1898}  &  yes      \\
Kohler            & K     & \cite{Kohler1957}                 &  yes      \\
Hudson-McCourbrey & HMC   & \cite{Hudson1960}                 &  yes      \\
Fender-Halsey     & FH    & \cite{Fender1962}                 &  yes      \\
Hiza              & H     & \cite{Hiza1969,Hiza1970,Hiza1978} &  no       \\
Sikora            & S     & \cite{Sikora1970}                 &  no       \\
Smith-Kong        & SK    & \cite{Smith1972,Kong1973}         &  no       \\
Halgren           & HHG   & \cite{Halgren1992}                &  no       \\
Waldman-Hagler    & WH    & \cite{Waldman1993}                &  no       \\
MATAR-1           & M1    & \cite{Al-Matar2004}               &  no       \\
MATAR-2           & M2    & \cite{Al-Matar2004}               &  no       \\\hline
\end{tabular}						      
\end{center}
\end{table}

\clearpage

\begin{table}[ht]
\noindent
\caption[]{Binary mixture types (1--6) classified according to the molecular models of the two components. \label{tab_mix_type}}
\bigskip
\begin{center}
\begin{tabular}{|l|c|c|c|} \hline
{}                   & 1CLJ & 2CLJQ & 2CLJD \\ \hline
1CLJ                 &   1  &   2   &  3    \\ 
2CLJQ                &      &   4   &  5    \\ 
2CLJD                &      &       &  6    \\ \hline
\end{tabular}
\end{center}
\end{table}

\clearpage

\begin{table}[ht]
\noindent
\caption[]{Unlike LJ energy parameters of the adjusted mixture models taken from \cite{Stoll2003,Vrabec2004,Stoll2004}, grouped according to the mixture type as defined by Table \ref{tab_mix_type}. \label{tab_uLJeps}}
\bigskip
\begin{center}
\begin{tabular}{|l@{ + }l|c|c||l@{ + }l|c|c|} \hline
\multicolumn{2}{|l|}{}&&&\multicolumn{2}{|l|}{}&& \\[-2.0ex]
\multicolumn{2}{|l|}{Mixture} & Type & $[\epsilon_{ab}^{\rm adj}/k_{\rm B}]$ / K& \multicolumn{2}{|l|}{Mixture} & Type & $[\epsilon_{ab}^{\rm adj}/k_{\rm B}]$ / K\\ 
\multicolumn{2}{|l|}{}&&&\multicolumn{2}{|l|}{}&& \\[-2.0ex]\hline
%\multicolumn{2}{|l|}{}& & K & \multicolumn{2}{|l|}{}& & K  \\ \hline
Ne & Ar         & 1 &\053.752 &           $\rm CO_2$   &$\rm C_2H_6$ &4& 128.878 \\
Ar & Kr         & 1 &136.280  & 	  $\rm CO_2 $  &$\rm CS_2 $  &4& 170.086 \\
Ar & $\rm CH_4$ & 1 &126.974  & 	  $\rm CO_2 $  &$\rm Cl_2 $  &4& 137.020 \\
Kr & Xe         & 1 &190.225  & 	  $\rm C_2H_4$ &$\rm C_2H_6$ &4& 106.470 \\
Kr & $\rm CH_4$ & 1 &156.184  & 	  $\rm C_2H_4$ &$\rm C_2H_2$ &4&\076.446 \\
Ne         & $\rm N_2$    &2&\032.513 &   $\rm C_2H_6$ &$\rm C_2H_2$ &4& 101.267 \\
Ne         & $\rm O_2$    &2&\035.249 &   $\rm C_2H_4$ &$\rm CO_2 $  &4&\095.579 \\
Ne         & $\rm CO_2$   &2&\075.559 &   $\rm C_2F_6$ &$\rm CO_2 $  &4& 105.045 \\
Ar         & $\rm N_2$    &2&\064.065 &   $\rm CO_2 $  & Propylene   &4& 130.390 \\
Ar         & $\rm O_2$    &2&\070.164 &   $\rm C_2H_6$ & Propylene   &4& 146.450 \\
Ar         & $\rm CO_2$   &2& 124.610 &   $\rm C_2H_4$ & Propylene   &4& 109.654 \\
Kr         & $\rm O_2$    &2&\082.030 &   $\rm N_2$    & Propylene   &4&\069.491 \\
$\rm CH_4$ & $\rm N_2$	  &2&\068.976 &   $\rm C_2H_6$ & CO  &5&\072.304\\
$\rm CH_4$ & $\rm CO_2$   &2& 135.331 &   $\rm CO_2 $  & CO  &5&\075.719\\
$\rm CH_4$ & $\rm C_2H_6$ &2& 142.225 &   $\rm N_2  $  & CO  &5&\036.134\\
$\rm CH_4$ & $\rm C_2H_4$ &2& 109.268 &   $\rm CS_2 $  & R22 &5& 203.132\\
$\rm CH_4$ & CO           &3&\074.256 &   $\rm CO_2 $  & R22 &5& 154.666\\
$\rm N_2 $   &$\rm O_2  $ &4&\039.091 &  $\rm CO_2 $  & R23 &5& 127.914\\
$\rm N_2 $   &$\rm CO_2  $&4&\070.979 &  R143a & R134a&6& 168.985  \\
$\rm N_2 $   &$\rm C_2H_6$&4&\067.344 &  R125 & R143a &6& 161.770  \\
$\rm N_2 $   &$\rm C_2H_4$&4&\049.851 &  R125 & R134a &6& 168.663  \\
$\rm O_2 $   &$\rm CO_2 $ &4&\074.255 &  R143a & R152a&6& 177.304  \\ \hline
\end{tabular}
\end{center}
\end{table}

\clearpage

\begin{table}[ht]
\noindent
\caption[]{Deviations of the unlike LJ size parameters from the arithmetic mean, i.e.~the Lorentz rule, for seven combining rules expressed by the root mean square ${\rm RMS}^{\rm x}_{\sigma}$ as defined by Equation (\ref {equ_rms_sigma}) for the different mixture types and for all 44 binary mixture models.}
\label{tab_rms_sig}
\bigskip
\begin{center}
\begin{tabular}{|l@{ Eq. }r|c|c|c|c|c|c|c|} \hline
\multicolumn{2}{|c|}{}                &\multicolumn{7}{|c|}{${\rm RMS}^{\rm x}_{\sigma}$ / \%} \\ \cline{3-9}
\multicolumn{2}{|c|}{Combining rule}  &\multicolumn{7}{|c|}{Mixture type}                      \\ \cline{3-9}
\multicolumn{2}{|c|}{}     & 1  & 2   & 3 & 4     & 5   & 6   & {\bf all}\\ \hline
S, & (\ref{equ_S_sig})     &0.0 & 0.0 & 0.0 & 0.0 & 0.0 & 0.0 & {\bf 0.0}\\
M2, & (\ref{equ_M2_sig})   &0.3 & 0.3 & 0.2 & 0.7 & 0.1 & 0.0 & {\bf 0.5}\\
SK, & (\ref{equ_SK_sig})   &0.7 & 0.6 & 1.1 & 0.5 & 0.3 & 0.0 & {\bf 0.6}\\
HHG, & (\ref{equ_HHG_sig}) &0.9 & 1.0 & 0.7 & 1.7 & 0.5 & 0.0 & {\bf 1.2}\\ 
H, & (\ref{equ_H_sig1})    &1.0 & 1.8 & 0.3 & 1.2 & 0.5 & 0.1 & {\bf 1.3}\\ 
WH, & (\ref{equ_WH_sig})   &1.0 & 1.2 & 0.9 & 2.0 & 0.7 & 0.0 & {\bf 1.4}\\ 
M1, & (\ref{equ_M1_sig})   &1.0 & 1.1 & 0.8 & 2.0 & 0.6 & 0.0 & {\bf 1.4}\\  \hline

\end{tabular}
\end{center}
\end{table}

%\clearpage

\begin{table}[ht]
\noindent
\caption[]{Deviations of the unlike LJ energy parameters from adjusted values for eleven combining rules expressed by the root mean square ${\rm RMS}^{\rm x}_{\epsilon}$ as defined by Equation (\ref {equ_rms}) for the different mixture types and for all 44 binary mixture models.  In the last column, it is indicated whether the unlike LJ size parameter is given by the arithmetic mean, i.e.~the Lorentz rule. \label{tab_rms_eps}}
\bigskip
\begin{center}
\begin{tabular}{|l@{ Eq. }r|c|c|c|c|c|c|c|c|} \hline
\multicolumn{2}{|c|}{}               & \multicolumn{7}{|c|}{${\rm RMS}^{\rm x}_{\epsilon}$ / \%} &            \\ \cline{3-9}
\multicolumn{2}{|c|}{Combining rule} & \multicolumn{7}{|c|}{Mixture type}                        &  {Lorentz} \\ \cline{3-9}
\multicolumn{2}{|c|}{}     & 1  & 2     &     3 &     4 &    5  &   6   & {\bf all   }   &  {}      \\ \hline
HMC, & (\ref{equ_HMC_eps}) & 4.9 & \04.9 & \01.6 & \03.1 & \04.3  & 1.4 & {\bf \03.9 } &  yes	  \\ 
M2, & (\ref{equ_M2_eps})   & 3.2 & \05.2 & \04.0 & \03.3 & \05.3  & 1.4 & {\bf \04.1 } &  no	  \\ 
SK, & (\ref{equ_SK_eps})   & 0.8 & \06.1 & \07.8 & \02.5 & \04.7  & 1.3 & {\bf \04.1 } &  no	  \\ 
S, & (\ref{equ_S_eps})     & 0.8 & \07.4 & \08.1 & \03.7 & \04.9  & 1.4 & {\bf \04.9 } &  no	  \\ 
LB, & (\ref{equ_LB_eps})   & 6.8 & \05.1 & \00.3 & \05.4 & \03.9  & 1.4 & {\bf \05.0 } &  yes	  \\ 
K, & (\ref{equ_K_eps2})    & 4.4 & 11.0  & \01.7 & \05.0 & \04.6  & 1.5 & {\bf \06.7 } &  yes	  \\ 
M1, & (\ref{equ_M1_eps})   & 0.7 & \07.2 & \06.2 & \08.2 & \06.7  & 1.4 & {\bf \06.8 } &  no	  \\ 
WH, & (\ref{equ_WH_eps})   & 0.8 & \07.4 & \06.6 & \08.0 & \07.0  & 1.4 & {\bf \06.8 } &  no	  \\ 
HHG, & (\ref{equ_HHG_eps}) & 2.9 & \09.1 & 11.5  & \06.8 & \08.8  & 1.4 & {\bf \07.3 } &  no	  \\ 
H, & (\ref{equ_H_eps1})    & 2.4 & 13.6  & \02.8 & \07.3 & \04.8  & 2.0 & {\bf \08.4 } &  no	  \\ 
FH, & (\ref{equ_FH_eps})   & 1.6 & 13.9  & 20.5  & 10.0  & \013.3 & 0.7 & {\bf 11.0  } &  yes	  \\ \hline

\end{tabular}
\end{center}
\end{table}
									    
\clearpage

% List of figures
\listoffigures
\clearpage

%\clearpage

%\clearpage

%\clearpage

%\clearpage

%\clearpage

%\clearpage

%\clearpage

% \begin{figure}[ht]
% \includegraphics[width=0.98\textwidth]{fig7}
% 
% \caption[Predicted unlike LJ energy parameters $\epsilon_{ab}^{\rm x}$ from eleven combining rules compared to adjusted values for the mixture types 2CLJQ+2CLJQ (4), 2CLJQ+2CLJD (5) and 2CLJD+2CLJD (6): {\Large $\circ$} LB, $\blacksquare$ K, {$\vartriangle$} HMC, $\blacktriangle$ FH, $\triangledown$ H, $\blacktriangledown$ S, $\square$ SK, {\Large $\bullet$} HHG, {\Large $\diamond$} WH, $\blacklozenge$ M1, + M2.]{Schnabel et al.\label{fig_Tpy4_5_6}}
% \end{figure}

\end{document}